\documentclass [12pt]{article}
\usepackage {graphicx}
\usepackage {longtable}

\begin{document}
\begin{center}
{\Large \textbf{Merger Driven Explosive Evolution of Distant Galaxies (Minor Mergers)}}
\end{center}

\medskip

\begin{center}
A.V.~Kats$^1$ and V.M.~Kontorovich$^{2,3}$
\end{center}

\begin{center}
\textit{\small{$^1$Institute of Radio Physics and Electronics,
National Academy of Sciences of Ukraine, Kharkov, 61085, Ukraine;}}\\

\textit{\small{$^2$Institute of Radio Astronomy, National Academy
of
Sciences of Ukraine, Kharkov, 61002, Ukraine;}}\\

\textit{\small{$^3$V.N.Karazin National University, Kharkov 61022, Ukraine;}}\\

\textit{\small{vkont1001@yahoo.com}}
\end{center}

\begin{abstract}
We derived solutions for the Smoluchowski
kinetic equation for the mass function of galaxies, which
describes mergers in differential approximation, where mergers
with low-mass galaxies are the dominant factor. The evolution of
the initial distribution is analyzed as well as the influence of
the source represented by galaxies (halos) that separate from the
global cosmological expansion. It is shown that the evolution of
the slope of the power-law part of the luminosity function at a
constant mass-to-luminosity ratio observed in the Ultra Deep
Hubble Field can be described as a result of explosive evolution
driven by galaxy mergers. In this case the exponent depends
exclusively on the uniformity degree of merger probability as a
function of mass.\\

pacs: /, 95.30.Sf, 98.35.Ce, 98.62.Ck, 98.80.-k, 95.35.+d\\

keywords /, galaxy: minor merging, mass function, evolution
\end{abstract}

DOI: 10.1134/S1990341313030012

\bigskip

\subsection*{INTRODUCTION}

In its generally adopted form the hierarchical clustering model
yields self-similar solutions \cite{b1,b2,b3}. At the same time,
investigations of distant galaxies reveal markedly
non-self-similar behavior, which shows up, among other things, in
the redshift dependence of the slope $\alpha $ of the power-law
part of the Schechter luminosity function (LF) \cite{b4,b5,b6}

\begin{equation}
\label{eq1} \phi (L) = \phi \,_{\ast}  \,L\,^{\alpha} \,exp\left(
{ - L/L\,_{\ast} }  \right)\,.
\end{equation}

\noindent (Here we do not discuss the parameters $\phi _{\ast}  $
and $L_{\ast}  $ of the Schechter function.)\footnote{At large
masses the MF is not exponential, but decreases in accordance with
a square-root law (Section 7), which is due to the adopted model
of source localization in the kinetic equation. } We show that the
observed evolution of the LF slope (Section 2), understood as the
slope of the mass function (MF) of galaxies, can be described as a
result of explosive evolution driven by galaxy mergers. Currently,
mergers are believed to be the factor responsible for the
evolution of types and masses of galaxies \cite{b7} (see also the
discussion and references in the reviews \cite{b8,b9,b10}).
Although the actual situation is more complex (see the recent
reviews by Ellis and Silk \cite{b11} and Silk and Mamon
\cite{b12}), we show that the observed parameters of the MF can be
explained satisfactorily in terms of our hypothesis. We derive
solutions for the Smoluchowski kinetic equation (KE) that
describes merger-driven explosive evolution of the mass function
of galaxies evolution of the mass function of galaxies
\cite{b13,b14} in differential approximation \cite{b13}, where
mergers of massive galaxies with low-mass galaxies (minor mergers)
play the main part (Sections 3 and 4). Note that the slope $\alpha
$ of the MF of massive galaxies, which is proportional to the
Schechter exponent $ \propto M^{\alpha} $, can be written
exclusively in terms of the uniformity degree $u$ of the merger
probability as a function of mass (Sections 5 and 6). This
property, in principle, makes it possible to determine the
probability of mergers as a function of mass including that of the
dark matter, based on observational data. In this paper we use the
well-known dependencies of the probabilities of galaxy mergers on
their mass to show that the observed evolution can be explained in
terms of natural assumptions about merger mechanisms in different
epochs (Sections 5 and 6). Explosive evolution occurs if the
exponent $u > 1$, which is evidently true for galaxy mergers. In
the case of explosive evolution of an initial MF of sufficiently
general form (i.e., decreasing faster than the square of the mass)
a power-law asymptotic develops with the exponent $\alpha = - u$;
in the case of predominant influence of a mass-localized source
the exponent is equal to $\alpha = - (u + 1)/2$ (Section 6).
Galaxies that separate from cosmological expansion play the role
of the source. At large red shifts $z$ = 6-8 the exponent $u$ is
determined by mergers of low-mass galaxies and is close to $u=2$.
At small $z$, where more massive galaxies merge, gravitational
focusing has to be taken into account \cite{b15} (see also
Appendix 4). At $z \to 0$, when using a radius-mass relation of
the form  $R \propto \sqrt {M} $, which follows from the
Tully-Fisher and Faber-Jackson relations\footnote{This exponent
proved to be redshift dependent (see references in
\cite{b11,b12}), but we do not take this fact into account in this
paper.}, this yields the well-known current value of the Schechter
index $\alpha = - 1.25$. At intermediate redshifts both the
evolution of the MF driven by the source (in the domain of low
masses) and the evolution of the initial mass distribution (in the
domain of sufficiently large masses) result in the Schechter slope
of $\alpha = - 1.5$, which can be reconciled with observational
data (Section 6). The above mechanisms of the formation of the MF
(with the allowance for the assumed contribution of dark matter)
alternate with each other in the process of evolution. Note that
in the $u = 2$ case we derived an exact solution of the kinetic
equation with a local source in a wide mass range. In the general
case of arbitrary $u > 1$ we derived the asymptotic form of the
solution of the kinetic equation at large masses on times close to
the moment of "explosion." This result can qualitatively explain
the observational data for the MF slope at the red shifts from $z
= 0$ to $z = 8$.

The MF derived in the approximation considered (Section 7) has a
form similar to that of the Schechter function, however, at large
masses it decreases in accordance with a square-root law and not
exponentially (see footnote~1). We determined the maximum mass
(Section 7) cutoff of the MF as a function of time and parameters
of the system in the vicinity of the "explosion" time. This
approach also makes it possible, in principle, to determine the
"explosion" times as a function of the initial conditions,
interaction parameters, and merger mechanism (see Appendix). Our
analysis is limited to the differential approximation and pairwise
mergers (with the allowance for the dependence of merger
probability on the masses of the galaxies involved). Despite these
restrictions, the inferred MF slopes for explosive solutions agree
satisfactorily with observations in a wide range of redshifts
\cite{b16}.

\subsection*{OBSERVATIONS OF GALAXIES \\AT LARGE REDSHIFTS}

The advances in observations of galaxies at large $z$ are largely
due to the use of multicolor photometry \cite{b17} in ultra deep
fields of major telescopes and observations of distant galaxies
through gravitational lenses. Let us mention, as a remarkable
example, the observations of a possible galaxy merger at $z = 2.9$
\cite{b18, b19}.

The methods of multicolor photometry, which isolate the galaxies
seen only through infrared filters because of the large
redshifts,made it possible to construct galaxy luminosity
functions out to z = 8 and even farther away based on decade-long
observations of ultra deep fields on the Hubble Space Telescope
and major ground-based telescopes. The Schechter LF slopes
$\alpha$ (Fig. 1) determined in a series of papers by Bouwens,
Illingworth, and their coauthors \cite{b4,b5} , which we use below
(see also numerous important references in
\cite{b6,b11,b12,b16,b20} ), depend significantly on redshift.
Note that the MF slope computed in hierarchical clustering models
without the allowance for mergers \cite{b1,b2,b3}  does not depend
on z.

\begin{figure}[h]
\includegraphics[scale=0.6]{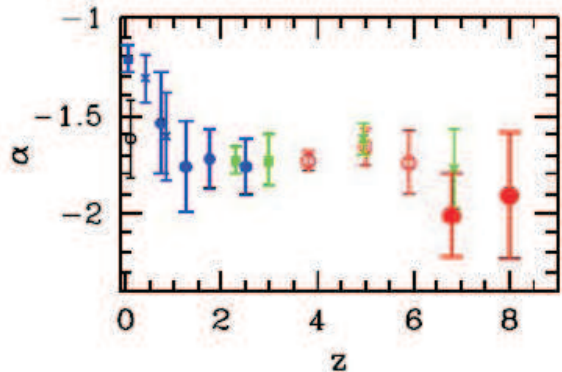}
\caption{Variation of the Schechter slope with redshift in HUDF
(the copy of Fig. 14 from \cite{b5} by courtesy of R.Bouwence). }
\end{figure}

\subsection*{SMOLUCHOWSKI KINETIC EQUATION \\DESCRIBING MERGERS}

We now consider the differential version of the Smoluchowski
kinetic equation for galaxy mergers with small mass increments
\cite{b7}. It has the following form (see Appendix 1):

\begin{equation}
\label{eq2} \frac{{\partial} }{{\partial t}}f(M,t) + C\Pi
\frac{{\partial }}{{\partial M}}\left[ {M^{u}f(M,t)} \right] =
\varphi (M,t),\quad \Pi = \int\limits_{0} {dM_2\, M_2 f(M_2,t)}\,.
\end{equation}

\noindent Here we use the expression for the probability of galaxy
mergers in the form $ÑM^{u}/2$ (see Appendix 2), where the
exponent $u > 1$. The $\Pi $ quantity is equal to the total mass
of low-mass galaxies ($M_{2} \ll M$), which is assumed to undergo
no significant changes in the merger process (below we adopt $\Pi
= Const$). The presence of the source term $\varphi $ in the
kinetic equation is essential; it describes the contribution of
massive galaxies separating from cosmological expansion as
gravitational instability develops \cite{b1,b2,b3}.

Let us rewrite the kinetic equation in the form

\begin{equation}
\label{eq3}
\begin{array}{l}
 \frac{{\partial} }{{\partial t}}F(M,t) + C\Pi
M^{u}\frac{\partial}{\partial M}F(M,t) = \Phi (M,t),\quad \\
 F(M,t) = M^{u}f(M,t),\quad \Phi (M,t) = M^{u}\varphi (M,t) \\
 \end{array}
\end{equation}

\noindent and use the method of characteristics to solve it. As a
result, the kinetic equation reduces to the following set of
ordinary differential equations

\begin{equation}
\label{eq4}
\begin{array}{l}
 {{dM} \mathord{\left/ {\vphantom {{dM} {dt}}} \right.
\kern-\nulldelimiterspace} {dt}} = C\Pi M^{u}\, \\
 {{dF} \mathord{\left/ {\vphantom {{dF} {dt = \Phi} }} \right.
\kern-\nulldelimiterspace} {dt = \Phi} } \\
 \end{array}.
\end{equation}

\noindent The solution of the kinetic equation is an arbitrary
function of independent first integrals of the equation set (4)
\cite{b21}. We find this function based on the initial conditions
(see Appendix 3). Integration of the first differential equation
of equation set (4) yields

\begin{equation}
\label{eq5} \tau (t) + \frac{1}{(u - 1)M^{u - 1}} = a(M,t) =
const,\quad \tau (t) \equiv C\int\limits_{0}^{t} {dt\,\Pi (t)} \to
C\Pi \cdot t.
\end{equation}

\noindent To integrate the second equation of the set, we first
assume that the source $\varphi $ is local. More specifically, we
assume that

\begin{equation}
\label{eq6} \Phi (M,t) = \delta (M - \bar {M}(t))\Phi (t),
\end{equation}

\noindent where $\Phi (t)$ is a function of time whose form is of
no importance to us. This assumption means that at time $t$ a
galaxy of mass $\bar {M}(t)$ separates from cosmological
expansion. We now write the mass $M$ in equation (6) in terms of
the first integral a derived above, as $M = \mu (a,t)$, where,
according to equation (5),

\begin{equation}
\label{eq7} \mu (a,t) = \left[ {(u - 1)\left( {a - \tau (t)}
\right)} \right]^{- \frac{1}{u - 1}}.
\end{equation}

\noindent This allows us to find the second independent integral
$b(M,t)$ of equation set (4) and construct the general solution of
the kinetic equation.

\subsection*{SOLUTION OF THE KINETIC EQUATION \\FOR A LOCALIZED SOURCE}

The kinetic equation (2) is linear, and therefore its solution has
the form of a sum of two independent terms

\begin{equation}
\label{eq8} f(M,t) = f_{in} (M,t) + f_s (M,t).
\end{equation}

\noindent The first term $f_{in} $ can be written in terms of the
initial distribution $f_{0} (M)$ (see Appendix) and contains only
the integral a. The second term $f_{s} $ is associated with the
source and contains the function $\bar {M}(t)$ that describes this
source and monotonically increases with time \footnote{The
explicit form of it, not playing the  fundamental role, we specify
below.}:

\begin{equation}
\label{eq9}
\begin{array}{l}
 f_{in} (M,t) = \left[ {(u - 1)\tau M^{u -
1} + 1} \right]^{ - \frac{u}{u - 1}}f_{0} \left\{ {M\left[ {\left(
{u - 1} \right)\tau M^{u - 1} + 1} \right]^{ - \,\frac{1}{u - 1}}}
\right\}\quad \\
 f_{s} (M,t) = M^{- u}K\left( {\tau + \frac{{1}}{{\left(
{u - 1} \right)M^{u - 1}}},t} \right) \\
 \end{array},
\end{equation}

\noindent where

\begin{equation}
\label{eq10}
\begin{array}{l}
 K(a,t) = \int\limits_{0}^{t} {dx\delta \left[ {\mu (a,x)
 - \bar {M}(x)} \right]\Phi (x)}= \\
 \sum\limits_{n} {\Phi (x_{n})\theta \left( {t - x_{n}}
\right)\left| {\frac{d}{dx}\left[ {\mu (a,x) - \bar
{M}(x)} \right]} \right|_{x = x_{n}} ^{- 1}}  \\
 \end{array}.
\end{equation}

Here $x_{n} $ -- are the roots of equation (11)

\begin{equation}
\label{eq11} \mu (a,x) = \bar {M}(x),
\end{equation}

\noindent which determines the zero points of the $\delta
$-function. It is important that $\bar {M}(t)$ is a monotonically
increasing function of time \cite{b1,b2,b3}. For simplicity, we
choose $\bar {M}(t)$ to be a linear function $M (t) = t/A$, and
can therefore construct for $u = 2$ an exact "explosive" solution
of the kinetic equation (see Appendix) and, for $u > 1$, find the
asymptotics in the domain of large masses near the "explosion"
time $t_{cr} $ that are of interest to us. The explosive behavior
of the solution \footnote{The explosive behavior of the solutions
of the Smoluchowski equation was first found by W. Stockmayer in
1943, and was then repeatedly rediscovered (see, e.g., references
in reviews \cite{b21,b22,b23,b24}).} at $u > 1$ found for galaxy
mergers by Kontorovich, Kats and Krivitskii \cite{b13} and
Cavaliere et al. \cite{b14} [14] means, as we shall see below,
that an initially localized mass distribution formally reaches the
domain of infinite masses in finite time (see \cite{b22,b23,b24}
).

\subsection*{EXACT SOLUTION IN THE CASE OF
QUADRATIC DEPENDENCE OF MERGER PROBABILITY ON MASS ($U = 2$)AND A SOURCE LINEAR IN TIME}

In this case, the equation for the roots of the arguments of the
$\delta$-function (11) becomes quadratic $ C\Pi \cdot x^{2} -
a\cdot x + A = 0 $ with the following roots

\begin{equation}
\label{eq12} x_{ \pm}  = \frac{{a \pm \sqrt {a^2 - a_{cr}^2} }
}{{2C\Pi }},\,\,\,a_{cr} \equiv 2\sqrt {AC\Pi} \, .
\end{equation}

\noindent A real solution exists for $a \ge a_{cr} $, and the
multiple root corresponds to the tangency of a hyperbola (into
which the left-hand side of equation (11) turns at $u = 2$,
according to equation (7)) and a straight line (the right-hand
side of the same equation, see Fig. 2). A multiple root (at time
$t_{tan} $ of the tangency of the straight line and hyperbola)
under the sign of the $\delta $-function is
unacceptable\footnote{To overcome this (model) restriction, we
apply a very simple regularization in the next section by
replacing the $\delta$-function with a step function of a small
finite width. }: at the critical value of parameter $a = a_{cr} $
the solution (10) becomes infinite. Below we overcome this
restriction. At the same time, small differences $a - a_{cr} $ are
of special importance to us, because they correspond to the case
of sufficiently large masses and time t close to the "explosion"
time $t_{cr} = 2t_{tan} $ that we are interested in (the domain of
large masses between the hyperbola $a = a_{cr} $ and the asymptote
of this hyperbola, $T \equiv t/t_{tan} = 2$, see Fig. 2).

\begin{figure*}[h]
\includegraphics[scale=1.2]{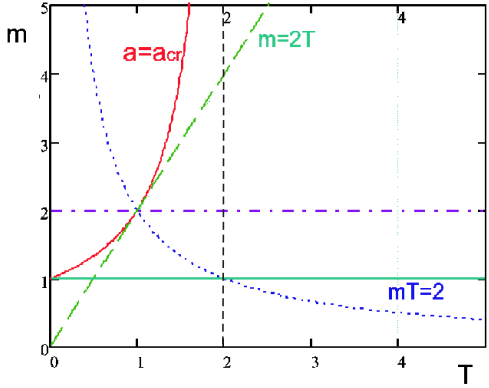}
\caption{\label{b}The domain of large masses $m \equiv a_{cr}M>2T,
2/T$ that is of interest to us lies between the hyperbola
$a=a_{cr}$ and its vertical asymptote (the dashed line) $T \equiv
{t}/{t_{tan}}=2$, which corresponds to the time of the explosion,
above the $m = 2T$ straight line. In this domain both roots of the
$\delta$-function contribute to the MF.  }
\end{figure*}

\noindent The contribution of the $\delta $-function of a complex
argument $\delta (y(x))$ to the integral is known to be given by
the following formula (the subscript numbers the roots of the
argument) $\delta (y(x)) = \sum\limits_{i} {\delta \left( {x_{i}}
\right)} \left| {\frac{dy}{dx}} \right|_{x_{i}} ^{ - 1} $.
If $y(x)$ is a quadratic trinomial, which we write in
the form $y(x) = \left( {x - x_{ +} } \right)\left( {x - x_{ -} }
\right)$, where $x_{ \pm} $ are the roots of the equation $y(x) =
0$, we derive, by virtue of $\frac{dy}{dx} = 2x - ( {x_{+} + x_{-}
} )$ the following two equations $\left. {\frac{dy}{dx}}
\right|_{x = x_{ +} } = x_{ +} - x_{ -} \,;\,\,\,\,\,\,\,\left.
{\frac{{dy}}{{dx}}} \right|_{x = x_{ -} } = - \left( {x_{+}  -
x_{-} }  \right)$.  It is evident that $\left| {\left. {\frac{dy}{dx}}
\right|_{x = x_{ \pm} } } \right| = \left| {x_{+}  - x_{-} }
\right| \propto \sqrt {a^2 - a_{cr}^2} $. The non-vanishing factor
under the sign of the $\delta $-function should not be differentiated,
because the corresponding extra terms vanish when the roots are
substituted. This simplifies the computations considerably. Thus
the solution with a $\delta $-function source and a quadratic
equation for its roots acquires the following form:

\begin{equation}
\label{eq13}
K\left( {a,t} \right) = \frac{{1}}{{\sqrt {a^{2} - a_{cr}^{2}}
}}\sum\limits_{ \pm}  {\tilde {\Phi} \left( {x_{ \pm} }  \right)\theta
\left( {a - a_{cr}}  \right)\theta \left( {t - x_{ \pm} }  \right)} ,
\end{equation}

\noindent where the tilde above $\Phi $ denotes the inclusion of
constant multipliers that arise in the form of coefficients at the
difference of the roots. We give the explicit formula in the
Appendix. For small $a - a_{cr} $  only this difference remains
under the square root sign in the denominator,

\begin{equation}
\label{eq14} K\left( {a,t} \right) \propto \frac{1}{{\sqrt {a -
a_{cr}} } }\,.
\end{equation}

\noindent Below we will make sure that in the general case the asymptotics for the solution at
large masses near the explosion time has a similar structure.

\subsection*{ POWER-LAW ASYMPTOTICS \\AT LARGE MASSES}

The above solution of the kinetic equation (9-11) can be used to
find the power-law part of the distribution of galaxy masses,
i.e., the exponent of the Schechter function at large masses in
the vicinity of the "explosion" time $t = t_{cr} $, under the
assumption of constant mass-to-luminosity ratio. Note that, as is
evident from the part of solution $f_{in} $ (9) that is determined
by the initial MF, the solution has the following asymptotics at
large masses (see Appendix 3)

\begin{equation}
\label{eq15} f_{in} \propto M^{- u}.
\end{equation}

\noindent The physical meaning of this part of the solution is
evident: it is a Kolmogorov-type distribution corresponding to a
constant flow of the number of massive galaxies\footnote{The
number of massive galaxies remains unchanged in the case of
massive galaxies merging with low-mass galaxies, which we consider
here. } along the mass spectrum ($U\left( {M,M_{2}}  \right)
\approx CM^{u}/2$):

\begin{equation}
\label{eq1} J = J\left( {M,t} \right) = 2f_{in} \int\limits_{0}
{dM_{2} M_{2} U\left({M,M_{2}}  \right) f_2 \,\rightarrow \,
f_{in}\left( {M,t} \right) CM^{u} \Pi } = \mbox{Const}.
\end{equation}

\noindent As for the part of the solution associated with the
source $f_{s} $, it is sufficient to consider the source in the
form of a $\delta$-function to determine the Schechter slope.
According to equation (14), an important feature of the
asymptotics of the derived solution for small $a - a_{cr} $
differences is the square root $\sqrt {a - a_{cr}}  $ in the
denominator. We are interested only in this asymptotics with the
substitution $a = a\left( {M,t} \right)$, which, according to
equation (5), corresponds to large masses and time instants close
to the explosion time, although in this case for $u = 2$ and a
linear right-hand part of equation (11) we can write the complete
solution. In view of $a\left( {M,t} \right) = M^{ - 1} + C\Pi t$
we have for $t \to t_{cr} \equiv a_{cr} /C\Pi $ the following
formula for $K\left( {a\left( {M,t} \right),t} \right)$:

\begin{equation}
\label{eq17}
K \propto \frac{{1}}{{\sqrt {a\left( {M,t} \right) - a_{cr}} } } =
\frac{{1}}{{\sqrt {\frac{{1}}{{M}} - C\Pi \left( {t_{cr} - t} \right)}} },
\end{equation}

\noindent where $t_{cr} $ corresponds to the vertical asymptote of
the hyperbola (7) for $a = a_{cr} $. Hence, given fs $f_{s}
\propto M^{ - u}K\left( {a,\,t} \right)$ it follows that for
sufficiently large masses and sufficiently small $t_{cr} - t$ from
(17) the dependence of the solution on mass has a power-law form

\begin{equation}
\label{eq18} f_{s} \propto M^{- 3/2}\,.
\end{equation}

In the general case $u \ne 2$ complex transcendent equations can
be derived for the coordinates of the tangent point and
intersection points of "hyperbola" (7) with the curves
corresponding to the right-hand side of equation (11). However, in
the asymptotic domain that is of interest to us, near the multiple
root of equation (11) and for small $\delta a \equiv a - a_{cr}
\ll a_{cr} $, we again have a quadratic equation. Therefore the
above reasoning remains true and the square root of $\delta a$
again appears in the denominator of the formula for $K(a,t)$ (see
Appendix 4):

\begin{equation}
\label{eq19} K\left( {a\left( {M,t} \right),\,t} \right) \propto
\frac{{1}}{{\sqrt {a\left( {M,t} \right) - a_{cr}} } } =
\frac{{1}}{{\sqrt {\left( {u - 1} \right)^{ - 1}M^{ - \left( {}^{u
- 1} \right)} - C\Pi \left( {t_{cr} - t} \right)}} }\,.
\end{equation}

\noindent It follows from this that although the position of the
tangent point and vertical asymptote of the hyperbola that
determines the position of the explosion point depends on all
parameters, the MF slope that we compute (the Schechter slope) is
determined solely by the dependence of the probability of mergers
on mass, i.e., only by its uniformity index $u$

\begin{equation}
\label{eq20} f_{s} = M^{ - u}K\left( {a\left( {M,t} \right),t}
\right) \propto M^{ - \frac{{u + 1}}{{2}}}\,.
\end{equation}

\noindent The uniformity index of the probability of galaxy
mergers is known for the two extreme cases \cite{b20,b25,b26}. If
the masses of merging galaxies are relatively small, then $u = 2$.
If the masses are sufficiently large, we must take into account
the gravitational focusing. In this case, the probability of
mergers is proportional to $ \propto \left( {M_{1} + M_{2}}
\right)\left( {R_{1} + R_{2}}  \right)$, where $R$ is the
characteristic radius of a galaxy.Hence the dependence of the
radius of a galaxy on its mass, $R \propto M^{\beta} $, becomes
important. The uniformity degree in the domain of large masses is
therefore equal to $u = 1 + \beta $. The Faber-Jackson and
Tully-Fisher relations imply $\beta = 1/2$. (The parameter $\beta
$ may differ from 1/2 at large $z$.)

We can thus make the following conclusions concerning the
Schechter slope $\alpha $ of the MF. We have $ - \alpha = u = 2$
in the domain of large $z$, where the galaxy masses are small and
the result is influenced by the initial MF. This slope coincides
with the results of observations at $z = 6-8$ (see Fig. 1). The
contribution of the source for small masses with $u = 2$ becomes
important at intermediate z, and we obtain $ - \alpha = \left( {u
+ 1} \right)/2 = 1.5$. At these $z$ the part of the distribution
that is determined by the initial MF is in the domain of
sufficiently large masses, where gravitational focusing should
manifest itself. This also results in $ - \alpha = u = 1 + \beta =
1.5$ and is approximately consistent with observations for $z =
3-5$. Finally, at small redshifts, where the source generates the
largest masses, $\alpha = \left( {u + 1} \right)/2 = 1 + \beta /2
= 1.25$, which is exactly equal to the well-known Schechter
exponent at the present epoch.

\subsection*{REGULARIZATION OF THE SOLUTION\\
AND MAXIMUM MASSES}

The solution with the $\delta$-function has an obvious drawback:
it is nonexistent (formally becomes infinite) in the case of a
multiple root in the argument of the $\delta $-function. Yet it is
the multiple root that corresponds to the maximum possible mass at
fixed $t$. The formally derived MF then goes to infinity. Hence
the MF that decreases in accordance with a power law at large
masses goes through a minimum whose position is easy to find in
the simplest cases. Thus at $u = 2$ the condition for a minimum is
satisfied at $M = M_{min} \left( {t} \right) =
\frac{{3}}{{4}}M_{max} \left( {t} \right)$, where $M_{max} \left(
{t} \right) = \frac{{1}}{{C\Pi \left( {t_{cr} - t} \right)}}$  is
the "maximum" mass\footnote{At $t \to t_{cr} $  this maximum mass
becomes infinite itself as a consequence of explosive evolution,
where infinite mass is attained in finite time.} at which $a =
a_{cr} $ and the derived MF becomes infinite.

\noindent The above means that the solution has to be regularized.
Regularization can be achieved by "spreading" the
$\delta$-function in one way or another. This procedure takes into
account the physically obvious condition that a galaxy cannot
instantly separates from cosmological expansion (because of the
finite buildup time scale of gravitational instability). This
condition can also be formulated in terms of masses assuming that
at a given time instant the galaxies that separates from
cosmological expansion lie within a small, but finite interval of
masses $\Delta \ll M$ near $\bar {M}\left( {t} \right)$. The
critical parameter $a_{cr} $ then splits into two parameters
$a_{cr} \to a_{cr}^{ \pm} $ ($a_{cr}^{ -}  > a_{cr}^{ +}  $)
corresponding to the tangent points of the hyperbola and the two
straight lines ($ \pm $) that bound the right-hand side of
equation (11), where (see Appendix)

\begin{equation}
\label{eq21} a_{cr}^{ \pm}  = \frac{{u}}{{u - 1}}\left( {AC\Pi}
\right)^{\frac{{u - 1}}{{u}}} \mp AC\Pi \frac{{\Delta} }{{2}},
\quad x^{ \pm}  = A\left( {AC\Pi}  \right)^{ - {{1}
\mathord{\left/ {\vphantom {{1} {u}}} \right.
\kern-\nulldelimiterspace} {u}}} \mp A\frac{{\Delta }}{{2}}\,.
\end{equation}

The regularized solution is finite and can be used to determine
the maximum mass $M_{max} \left( {t} \right)$ of the distribution
as a function of time (we restrict this analysis to proper time).

\begin{figure*}[h]
\includegraphics[scale=1.4]{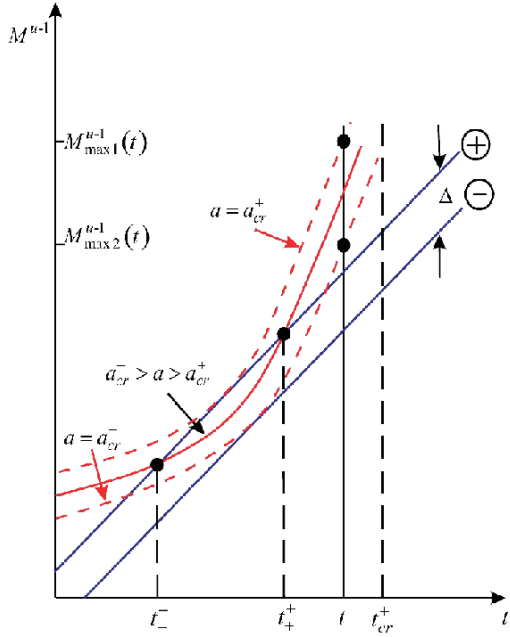}
\caption{The case of two intersection points of the hyperbola and
the ($ \pm $) lines bounding the nonzero source in the kinetic
equation at $a_{cr}^{ -}> a > a_{cr}^{ +} $. The $a = a_{cr}^{
\pm}  $ values of the integral correspond to the tangent points of
the hyperbola and ($ \pm $) lines. The times $t_{ \pm} ^{ +}  $
correspond to the intersection of the hyperbola with the upper
line (+) at $a > a_{cr}^{ -}  $. This case does not allow the
transition to $\Delta = 0$. This mass domain corresponds to the
largest MF masses lying in the interval $M_{max2}(t) < M <$
$M_{max1} (t)$\,.  }
\end{figure*}

Let us assume for simplicity $\Phi \left( {t} \right) = \Phi $.
Only those portions where the right-hand side of equation (11)
differs from zero contribute to the integral K (10). The solution
of the kinetic equation with a source in the domain of parameters
satisfying the condition $a > a_{cr}^{ -}  > a_{cr}^{ +}  $ has
the following form for $t > t_{ +} ^{ +}  $ (see Appendix 5 and
Fig. 3)

\begin{equation}
\label{eq22}
K\left( {a,t} \right) = \frac{{2\Phi \varsigma} }{{\Delta} }\{ \sqrt {a -
a_{cr}^{ +} }  - \sqrt {a - a_{cr}^{ -} }  \} ,
\end{equation}

\noindent
where $\varsigma = \sqrt {\left( { - d\mu /da}
\right)/\left( {d^{2}\mu /dt^{2}} \right)} $  at $a = a_{cr} $. We
derive from this the same result at $\Delta \to 0$ with a feature
corresponding to the source in the form of the $\delta$-function:

\begin{equation}
\label{eq23} K\left( {a,t} \right) \to \Phi \varsigma
\frac{{AC\Pi} }{{\sqrt {a - a_{cr} }} }\,.
\end{equation}

\noindent
For $\Delta \ne 0$

\begin{equation}
\label{eq24} K\left( {a,t} \right) = \frac{{2\varsigma \,\Phi}
}{{\Delta }}\frac{{a_{cr}^{ -}  - a_{cr}^{ +} } }{{\sqrt {a -
a_{cr}^{ +} }  + \sqrt {a - a_{cr}^{ -} } } } = \frac{{2\varsigma
\,\Phi \cdot AC\Pi} }{{\sqrt {a - a_{cr}^{ +} }  + \sqrt {a -
a_{cr}^{ -} } } }\,.
\end{equation}

\noindent
For $a \to a_{cr}^{ -}  $ cr the mass and the
corresponding MF value reach their maxima for this domain
 $t_{cr}^{ - } > t > t_{ -} ^{ +}  $
($t_{cr}^{ \pm}  \equiv a_{cr}^{ \pm}  /C\Pi $)

\begin{equation}
\label{eq25} K_{max} = \frac{{2\varsigma \,\Phi \sqrt {AC\Pi} }
}{{\sqrt {\Delta} }},\,\,\,M_{max}^{ -}  \left( {t} \right) =
\frac{{1}}{{\left[ {\left( {u - 1} \right)C\Pi \cdot \left(
{t_{cr}^{ -}  - t} \right)} \right]^{\frac{{1}}{{u - 1}}}}}\,.
\end{equation}

\noindent Thus in the case of a finite spread $\Delta $ we obtain
a finite result for the MF at the point corresponding to the
maximum mass for the domain considered. This MF value depends on
the spread $\Delta $, which becomes a measurable physical
parameter.

Consider now the domain of parameters corresponding to even larger
masses, i.e., to the case of two intersections $a_{cr}^{ -}  > a >
a_{cr}^{ +}  $ (see Fig. 4). For $t > t_{ +} ^{ +}  $

\begin{equation}
\label{eq26} K = \frac{{2\varsigma \cdot \Phi} }{{\Delta} }\sqrt
{a - a_{cr}^{ +} } \, .
\end{equation}

\noindent The largest mass $M_{max1}(t)$ corresponds to the upper
boundary $a = a_{cr}^{ +}  $  of the domain where the MF becomes
zero. As is evident from equation (26), K vanishes in accordance
with a square-root law. This dependence is superimposed by a
power-law decrease due to the relation $f = M^{ - u}F\left( {M,t}
\right)$. We thus derive a Schechter function analog for the MF,
which differs in the decrease law at large masses: the exponential
decrease described by equation (1) is replaced by the square-root
decrease $\sqrt {M_{max1} - M} $. Such a behavior, however, is
determined by the form of the source localization, which vanishes
outside the band $\Delta $. The maximum mass is equal to

\begin{equation}
\label{eq27} M_{max1} \left( {t} \right) = \frac{1}{{\left[
{\left( {u - 1} \right)C\Pi \left( {t_{cr}^{+} - t} \right)}
\right]^{\frac{{1}}{u - 1}}}}\,.
\end{equation}

\noindent All the above formulas describe explicitly the explosive
behavior of the MF. At $t \to t_{cr}^{ +}  - 0$ the maximum mass
goes to infinity.

\begin{figure*}[h]
\includegraphics[scale=1.4]{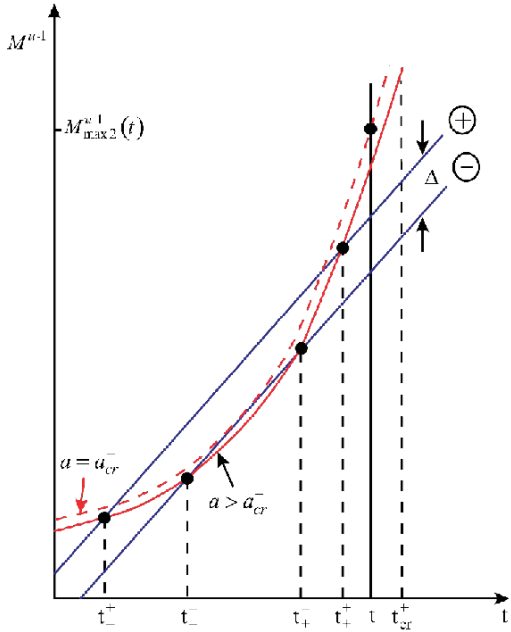}
\caption{The case of four points of intersection between the
hyperbola and the ($ \pm $) lines bounding a nonzero source in the
kinetic equation $a>a_{cr}^{-}$. Here $\Delta$ is the mass spread
width of the localized source. The value of the integral
$a=a_{cr}^{-}$ corresponds to the tangent point of the hyperbola
and the lower line. The times $t_{\pm}^{+}$ correspond to the
intersection of the hyperbola with the upper (+) and lower (-)
bounding lines, respectively, at $a>a_{cr}^{-}$. The case
considered allows the transition to $ \Delta =0 $. The position of
the maximum mass $ M_{max2} $ in this mass domain is indicated.}
\end{figure*}

We derived a solution in the form of a modified Schechter function
(Fig. 5) with a power-law portion at the "light" end and
square-root vanishing at the "heavy" end. Unlike the estimated MF
slope (Section 6), this result is model dependent. The very
presence of a local maximum in the domain of large masses appears
to indicate that the spread parameter $\Delta $ is sufficiently
small, and requires further investigation. Note that if considered as a function of parameter $a$, the MF
decreases as $1/\sqrt {a} $, making the domain of the smallest
possible parameter values that we analyzed above especially
important.

\subsection*{CONCLUSIONS}

In this paper we restricted our analysis to the
discussion of the slope (the Schechter slope) of the mass function
of galaxies, which is most simply related to the physical
properties of the problem in terms of the merger model: it is
determined solely by the exponent of the dependence of the cross
section of galaxy mergers on the mass of the biggest galaxy. The
solution obtained confirms that explosive evolution of the MF as a
result of minor mergers may, in principle, explain the observed
evolution of the low-mass end of the MF up to $z = 8$. The
computation of the other MF parameters requires invoking a
considerably greater amount of astronomical
data.\footnote{Furthermore, in this case we have to face a number
of unsolved problems, e.g., the so-called downsizing, where
starting from $z\approx 2-3$ the luminosities decrease toward z =
0 instead of continuing to increase. A possible solution of this
paradox may consist of taking into account the contribution of
galactic activity to luminosity.} The most important requirement
for the obtained solution is that the time scale of explosive
evolution should be shorter than the Hubble time. This criterion
can be satisfied only by imposing certain constraints on the
masses, radii, and velocities of galaxies with the allowance for
the dominating contribution of dark matter \cite{b6,b15}. In
particular, the average density of the mass contained in galaxies
(halos) should be more than two orders of magnitude higher than
the average density of matter in the Universe. Thus the mergers
seem to occur in groups and protoclusters inside larger-scale
halos and also in the walls, filaments, and knots of the
large-scale structure \cite{b27}. The interaction in the process
of merging is purely gravitational and we therefore used the known
galaxy merger probabilities in our estimates \cite{b20,b25,b26}. A
proper description of observational data may mean that dark matter
behaves as a collisional medium in the process of merging, as it
was already pointed out in the literature (see references in
\cite{b11} ). This may be a result of violent relaxation
\cite{b28}  driven by strong fluctuations of the gravitational
field in galaxy mergers.

\begin{figure*}[h]
\includegraphics[scale=1.6]{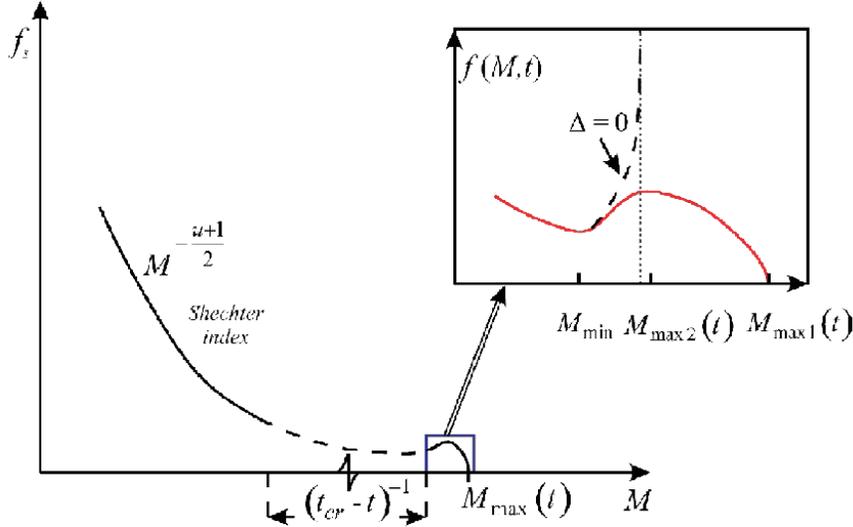}
\caption{The mass function $f_{s} = M^{ - u}K\left( {M,t} \right)$
established as a result of mergers with small mass increments. The
dashed line shows the MF singularity in the case of a
$\delta$-function source. The MF vanishes at $M=M_{max}(t)$
because of the adopted stepwise regularization model}
\end{figure*}

Note a characteristic difference between
the form of the MF derived and the Schechter function: a rising
portion of the MF superimposed on the overall decline before the
maximum masses (Fig. 5). The parameters of this rising portion are
related to the scale lengths of gravitational instability and
collapse during the separation of a galaxy from general
cosmological expansion.

\subsection*{ACKNOWLEDGMENTS}

We are grateful to the participants of the "Astroparticle Physics"
(Kiev, 2011), "Gravitation and Cosmology" (Kiev, 2012),
"Electromagnetic Methods of Environmental Studies" (Kharkov,
2012), and "Topical problems of extragalactic astronomy"
(Pushchino, 2013) conferences for useful critical comments, and to
A. Donets and Yu. Rudnev for their assistance with preparing the
figures. This work was supported in part by the
"Cosmomicrophysics" program of the National Academy of Sciences of
Ukraine.

\subsection*{APPENDIX}

\textbf{A1. Derivation of the Differential Kinetic Equation}

The Smoluchowski integral kinetic equation \cite{b20}

\[
\frac{\partial}{\partial t}f(M,t) = I_{st} + \varphi (M,t)
\]

\noindent with collision integral

\[
I_{st} = \int\!\!\!\int {dM_{1}}  dM_2 \left\{ {W_{M|M_1 , M_2}
f_1 f_2 - W_{M_1 |M_2 ,M} f_2 f - W_{M_2 |M,M_{1}} ff_1} \right\},
\]

\[
W_{M|M_1, M_2}  = U(M_1, M_2)\delta (M - M_1 -
M_2),\]
\[U ( M_2, M_1 ) = U( M_1, M_2 ),\quad f = f(M,t),\quad
etc.,\]

\noindent can be reduced to a differential equation if the kernel
function $U\left( {M_{1} ,M_{2}} \right)$ is not singular at
$M_{2} \to 0$, which is the case for galaxy mergers \cite{b20,b25}

\[
U\left( {M_{1} ,M_{2}}  \right) \to \frac{{C}}{{2}}M_{^{1}}^{u} \quad ,\quad
M_{1} \gg M_{2} ,
\]

\noindent if the main contribution to $I_{st} $ is provided by
mergers of massive galaxies with low-mass galaxies. In this case
the integrand can be expanded into $M_{2} \ll M,M_{1} $ (see
\cite{b9,b10}) and reduced to the form (2), where $\Pi $ is the
total mass of low-mass galaxies. We now rewrite $I_{st} $ in the form

\[
I_{st} = 2\int \limits_{0}^{M/2} \! dM_2\,
\{ U(M-M_2, M_2)\,f_2 f(M-M_2,t)-U(M_1, M_2)f_2 f \}
\]
\[-2\int\limits_{M/2}^{\infty}\! dM_2\, U(M,
M_2)\, f_2 f\,.
\]

\noindent We then expand the functions containing the difference
$M - M_{2} $ into $M_{2} \ll {{M} \mathord{\left/ {\vphantom {{M}
{2}}} \right. \kern-\nulldelimiterspace} {2}}$ and drop the terms
that are small in the ratio of small to large masses to reduce the
collision integral to the form $I_{st} \simeq - \frac{{\partial
J}}{{\partial M}},$ where $J$ denotes the flux of the number of
massive galaxies across the spectrum. Under the assumptions made
it can be written in the form

$$J(M,t) = CM^{u}f(M,t)\Pi (t),$$
\hspace{12cm} (A.1)
$$\Pi(t) = \int\limits_{0}^{M/2}
dM_2 \, M_2 f(M_2,t)\,.  $$

\noindent We drop the upper limit, because we assume that the main
contribution to $\Pi $ is provided by masses $M_{2} \ll M$. Note
that the Smoluchowski equation is derived from the full kinetic
equation for the distribution function that depends not only on
masses, but also on the coordinates and velocities (momenta) of
galaxies via an averaging procedure. Long-range interactions
result in faster establishment of velocity distributions compared
to mass distributions and therefore the distribution function can
be factorized. We now average it over the velocities and, given
the linear nature of the equation in the approximation of minor
mergers, derive an equation of the form (2), where the coefficient
C? appears under the averaging sign. In the case of a sufficiently
narrow velocity distribution this results in replacing the
velocities with their mean values. Averaging over the coordinates
results in the appearance of the mean value with the weight
accounting for the large-scale structure, which requires a special
analysis.

\textbf{A2. Probability of Galaxy Mergers}

The coefficient in the mass conservation law in the merger
probability $U\left( {M_{1} ,M_{2}}  \right) \to U = < \sigma v >
$, where the angular brackets, which we hereafter omit, denote
averaging over momenta $p$, $\sigma $ is the merger cross section,
and $v = |v_{1} - v_{2}|$ is the relative velocity of the two
galaxies. The cross section is equal to $\sigma = \pi R^{2}\left(
{1 + GM/Rv^{2}} \right)\varphi $, where $GM/Rv^{2}$ is the
gravitational focusing parameter \cite{b15} [15]; $\varphi =
\left( {1 + Rv^{2}/GM} \right)^{ - \xi} $ is the multiplier that
accounts for the dependence of merger probability in the case of a
head-on collision on relative velocity. In these formulas we adopt
$\xi > 0$, $R = R_{1} + R_{2} $,  and $M = M_{1} + M_{2} $. At
$GM/Rv^{2} \gg 1$ the formula for the cross section becomes
$\sigma \approx RM \cdot \pi G/v^{2}$. The probability uniformity
degree is determined by the parameter $u$, which is determined by
expression $U\left( {\lambda M_{1} \,,\lambda M_{2}} \right) =
\lambda ^{u}U\left( {M_{1} \,,M_{2}}  \right)$. The dependence of
the radius on mass, $R_{1} \propto M_{1}^{\beta}  $ , is
determined by the Tully-Fisher and Faber-Jackson laws
\cite{b1,b11}. The cross section may, in the general case, contain
dimensionless multipliers depending on the ratios of masses,
radii, and velocities, which have no effect on the uniformity
degree, but are nevertheless important for the asymptotics. Here
we adopt the simplest "elastic" variant \cite{b25,b26} for a
collisionless velocity distribution, for which

\[
u = 1 + \beta ,
\quad
\left( {GM/Rv^{2} \gg 1} \right);
\quad
u = \xi \left( {1 - \beta}  \right) + 2\beta ,
\]
\[\left( {GM/Rv^{2} \ll 1,\,\,\xi \le 2} \right);
\quad u = 2,\,\,( \xi \ge 2 )\,.\]

As for the order of magnitude of the probability and,
consequently, the estimated time scale of explosive evolution,
there still remains a large degree of uncertainty, even without
taking into account the so far insufficiently known properties of
dark matter. However, we assume that this time scale is shorter
than the Hubble time. For this to be true, the condition $\Pi
/\rho \ge 10^{2}$, at least, should be satisfied, where $\rho $ is
the average matter density in the Universe. According to
\cite{b3}, such a condition is satisfied in the case of halo
collapse.

\textbf{A3. Solving the Kinetic Equation Using the Method of
Characteristics}

For convenience, we denote the mass that appears in the first
integral at time $t = 0$ as $M_{0} $. We then have
$$a( M_0, t = 0) =
\frac{1}{{(u - 1)M_0^{u - 1}}}; M_{0} = \left[ {\frac{1}{{(u -
1)a( t = 0)}}} \right]^{\frac{1}{u - 1}};$$ $$f_{0} (M_0) = f_{0}
\left( {\left[ {\frac{1}{{(u - 1)a(t = 0)}}} \right]^{\frac{1}{u -
1}}} \right)\,.$$

Accordingly, $F(M,t) = M^{u}f(M,t)$; $F(M_{0} ,t = 0) = M_{0}
^{u}f_{0} (M_{0})$; whence it follows that
$$
F(M_{0}, t = 0) = \left[ {\frac{1}{{( {u - 1})a\left( {t = 0}
\right)}}} \right]^{\frac{{u}}{{u - 1}}}f_{0}
\left( {\left[ {\frac{{1}}{{\left( {u - 1} \right)a\left( {t = 0}
\right)}}} \right]^{\frac{{1}}{{u - 1}}}} \right).
$$
It follows
from this for arbitrary $M$ and $t$ that
$$
F\left( {M,t} \right) = \left[ {\frac{{1}}{{\left( {u - 1}
\right)a\left( {M,t} \right)}}} \right]^{\frac{{u}}{{u - 1}}}f_{0}
\left( {\left[ {\frac{{1}}{{\left( {u - 1} \right)a\left( {M,t}
\right)}}} \right]^{\frac{{1}}{{u - 1}}}} \right)\,.
$$
As it must be, $F\left(
{M,t} \right)
$ is a function of integral $a\left( {M,t} \right)$,
which satisfies the initial condition. We now substitute $a(M,t)$
in the form

$$\frac{{1}}{{\left( {u - 1} \right)a\left( {M,t} \right)}} = \frac{{M^{u -
1}}}{{C\Pi t\left( {u - 1} \right)M^{u - 1} + 1}}$$ and use
$f(M,t) = M^{- u}F(M,t)$ to derive formula  (9) for $f_{in}
(M,t)$. The first integral $b(M,t)$ of the second equation in (4)
for the zero initial condition for this part of the solution
results in formula (9) for $f_{s} (M,t)\,.$

\textbf{A4. Roots of the Delta-Function} Let us now consider the
more general case of \textbf{} $u > 1$, $\bar {M}\left( {t}
\right) = {{t^{s}} \mathord{\left/ {\vphantom {{t^{s}} {A}}}
\right. \kern-\nulldelimiterspace} {A}}$, $\Pi = const$. For the
above conditions

\[
\tau \left( {x} \right) = C\Pi \cdot x,\quad \mu \left( {a,x} \right) =
\left[ {\left( {u - 1} \right)\left( {a - C\Pi x} \right)}
\right]^{\frac{{1}}{{1 - u}}},
\]

\noindent and the equation for the roots of delta-function
acquires the following form

$$\frac{{1}}{{a - C\Pi x}} = \frac{{u - 1}}{{A^{u - 1}}}x^{\left( {u - 1}
\right)s}\,. \hspace {5cm}(A.2)$$

\noindent The left-hand part of the equation as a function of $x$
is a (generalized) hyperbola $(a > 0)$ with the vertical asymptote
$x_{as} = a\left( {C\Pi} \right)^{ - 1}$, and the right-hand part
is a growing power-law function. Equation (A2) may have no real
roots (if the hyperbola does not intersect with the power-law
function), or have two different roots, or one multiple root in
the case if the two curves touch each other. The condition to be
satisfied in the latter case is that the two functions and their
derivatives should be equal at the tangent point. It is more
convenient to use the logarithmic derivative

$$\frac{{1}}{{a - C\Pi x}} = \left( {u - 1} \right)sx^{-1}\,.\hspace {5cm}(A.3)$$

\noindent We now exclude $a - C\Pi x$ from formulas above
to find the tangent point

$$x_{tan}^{\left( {u - 1} \right)s + 1} = \frac{{s}}{{C\Pi} }A^{u-1}\,.\hspace {6cm}(A.4)$$

\noindent We then substitute $x_{tan} $ to find the value of
parameter $a$ corresponding to the multiple root:

$$a = a_{cr} = C\Pi \left[ {1 + \frac{{1}}{{\left( {u - 1} \right)s}}}
\right]x_{tan}=$$ $$\hspace {1cm}= C\Pi \left[ {1 +
\frac{{1}}{{\left( {u - 1} \right)s}}} \right]\left[
{\frac{{s}}{{C\Pi} }A^{u - 1}} \right]^{\frac{{1}}{{\left( {u - 1}
\right)s + 1}}}\,.\hspace {1cm}(A.5)$$

In the $s = 1$ case the formulas simplify to

$$a_{cr} = \frac{{u}}{{u - 1}}\left( {AC\Pi}  \right)^{1 - \frac{{1}}{{u}}},
\quad
x_{tan}^{u} = \frac{{1}}{{C\Pi} }A^{u - 1}.\hspace {3cm}(A.6)$$

\noindent In the simplest case of $u = 2$ and $s = 1$ they yield

$$a_{cr} = 2\sqrt {AC\Pi}  ,\quad x_{tan}^{} = \sqrt {\frac{{A}}{{C\Pi} }}\,.\hspace {5cm}(A.7)
$$

\noindent Thus for $a > a_{cr} $ we have two roots:  $x_{ -}  <
x_{tan} $, $x_{ +}  > x_{tan} $.

\noindent We now return to the general case to obtain, by
introducing parameter $\gamma = \left( {u - 1} \right)s$,

$$x_{tan}^{\gamma + 1} = \frac{{s}}{{C\Pi} }A^{u - 1},
\quad a_{cr} = C\Pi \left[ {1 + \frac{{1}}{{\gamma} }}
\right]\left[ {\frac{{s}}{{C\Pi} }A^{u - 1}}
\right]^{\frac{{1}}{{\gamma + 1}}}\,.\hspace {1cm}(A.8)$$

\noindent The master equation in dimensionless variables $\tilde
{a} = a/a_{cr} ,\quad T = x/x_{tan} $ has the form:

$$\frac{{1}}{{\left( {\gamma + 1} \right)\tilde {a} - \gamma T}} = T^{\gamma
}\,.\hspace {6cm}(A.9)$$

\noindent Expanding it into small $0 < \tilde {a} - 1 \ll 1$ and
$\left| {\delta T} \right| \ll 1$ in the vicinity of the tangent
point yields the quadratic equation
$
\gamma \left( {\delta T}
\right)^{2} - 2\gamma \left( {\tilde {a} - 1} \right)\delta T -
2\left( {\tilde {a} - 1} \right) = 0, $ roots at $\tilde {a}
- 1 \ll 1
$ are equal to
$$\delta T_{ \pm}  \approx \pm \sqrt {2\left( {\tilde {a} - 1} \right)/\gamma
}\, .\hspace {6cm}(A.10)$$

\textbf{A5. Regularization } We now replace the $\delta $-function
in the right-hand part of the kinetic equation by a step function
of finite width $\Delta > 0$

$$\Phi (M,t) = \delta _{\Pi}  (M - \bar {M}(t))
\Phi (t), $$ $$\quad \quad\delta _{\Pi} (x)= \frac{1}{{\Delta}
}\left[ {\theta \left( {x + \frac{{\Delta} }{2}} \right) - \theta
\left( {x - \frac{{\Delta} }{{2}}} \right)} \right]\Phi (t),\\
\hspace {2cm}(A.11)$$

\noindent where $\theta (x)$ is the Heaviside function. $\theta
(x) = 0$ for $x < 0$ and $\theta (x) = 1$ for $x \ge 0$. We then
have for $K(a,t)$:

$$\quad K(a,t) = \int\limits_{0}^{t} {dx\delta _{\Pi}  \left[ {\mu
(a,x) - \bar {M}(x)} \right]\Phi (x)}\,.\hspace {3cm}(A.12)$$

\noindent The integrand differs from zero if

$$\bar {M}(x) - \frac{\Delta}{2} \le \mu (a,x)
\le \bar {M}(x) + \frac{{\Delta} }{2}\,. \hspace {5cm}(A.13)$$

\noindent Each of the two boundary lines $y = \bar {M}(x) + \sigma
\frac{{\Delta} }{2}$ (where $\sigma = \pm $) may intersect twice
the hyperbola
$y = \mu (a,x)$ at $a > a_{cr}^{\sigma} $. The case $a = a_{cr}^{\sigma}$
corresponds to the contact of the curve
$ \bar {M}(x) + \sigma \Delta /2$   and hyperbola
$\mu(a,t)$
at the points $x = x^{\sigma} $ found from the following
conditions:

\hspace {11cm}(A.14)
\begin{eqnarray*}
{\mu (a,x)=\bar M(x)+ \sigma \frac{\Delta}{2} }\\
\quad \quad \quad  {\frac{{\partial \mu (a,x)}}{{\partial x}}=
\frac{d\bar M(x)}{dt} }\,.
\end{eqnarray*}

\noindent We find at $\Pi = const,\quad \bar {M}(t) = x/A$,

$$\frac{x^{\sigma} }{A} = \left( {AC\Pi}  \right)^{ - {1
\mathord{\left/ {\vphantom {{1} {u}}} \right.
\kern-\nulldelimiterspace} {u}}} - \sigma \frac{{\Delta} }{{2}},$$
$$\quad a_{cr}^{\sigma}  = \frac{{u}}{{u - 1}}\left( {AC\Pi}
\right)^{\frac{{u - 1}}{{u}}} - \sigma AC\Pi \frac{{\Delta
}}{{2}},\quad \sigma = \pm \,. \hspace {3cm}(A.15)$$

\noindent Note that  $x^{ +}  < x^{ -} $ and $a_{cr}^{ +}  <
a_{cr}^{ -} $, and $a_{cr}^{ -}  - a_{cr}^{ +}  = AC\Pi \cdot
\Delta \,.$ We obtain two points of intersection $x_{ -} ^{ +}
\left( {a} \right) \le x_{ + }^{ +}  \left( {a} \right)$ for
$a_{cr}^{ +}  < a < a_{cr}^{ -}  $ (fig.3) and four points of
intersection $x_{ \pm} ^{ \pm} \left( {a} \right)$(with both
critical lines) for $a_{cr}^{ -}  < a$ (fig.4). In the latter case
we can pass to the limit of $\Delta \to 0$ to get the $\delta
$-function in the righthand part of the kinetic equation:

$$ \sqrt {a-a_{cr}^{\sigma}} \approx \sqrt {a-a_{cr}}\left(
1+\sigma \frac{\Delta}{4} \frac{AC\Pi}{(a-a_{cr})} \right), $$
whence it follows that (cf. (22)) $$ \sqrt {a-a_{cr}^{-}} -\sqrt
{a-a_{cr}^{+}}\approx \frac{\Delta}{2} \frac{AC\Pi}{(a-a_{cr})}\,.
$$

In the case of small deviations of intersection points from the
tangent points we have $\delta x_{ \pm }^{\sigma}  \equiv x_{ \pm}
^{\sigma}  - x^{\sigma} $, where we substituted  $x_{ \pm} \left(
{a} \right)$ into $a = a_{cr}^{\sigma} $, as follows from equation
above, $\delta x_{ +} ^{\sigma} = - \delta x_{ -} ^{\sigma}  >
0\,.$ For $t > x_{ +} ^{ +}  $

$$K = K^{ +}  \equiv \frac{{1}}{{\Delta} }\int\limits_{t_{ -} ^{ +} } ^{t_{ +
}^{ +} }  {x\Phi \left( {x} \right)} \to \frac{{\Phi} }{{\Delta}
}\left( {t_{ +} ^{ +}  - t_{ -} ^{ +} }  \right) \approx
\frac{{\Phi} }{{\Delta }}\left( {\delta t_{ +} ^{ +}  - \delta t_{
-} ^{ +} }  \right) = \frac{{2\Phi} }{{\Delta} }\delta t_{ +} ^{
+} \, .$$

At $a_{cr}^{ -}  < a$ è $t > x_{ +} ^{ +}   \quad K\left( {a,t}
\right) = K^{ + } + K^{ -} $, where $$K^{ -}  = \frac{{\Phi}
}{{\Delta} }\left( {\delta t_{ + }^{ -}  - \delta t_{ -} ^{ -} }
\right) = \frac{{2\Phi} }{{\Delta} }\delta t_{ +} ^{ -} \,.
\hspace {5cm}(A.16)$$

\noindent Correspondingly,

$$K\left( {a,t} \right) = \frac{{\Phi} }{{\Delta} }\left( {\delta t_{ -} ^{ +
} - \delta t_{ +} ^{ +}  + \delta t_{ +} ^{ +}  - \delta t_{ +} ^{ -} }
\right) =$$ $$= \frac{{2\Phi} }{{\Delta} }\left( {\delta t_{ -} ^{ +}  + \delta
t_{ +} ^{ +} }  \right) = \frac{{2\Phi \varsigma} }{{\Delta} }\left( {\sqrt
{a - a_{cr}^{ +} }  - \sqrt {a - a_{cr}^{ -} } }  \right),\hspace {2cm}(A.17)$$

\noindent where we took into account that  $\delta t_{ \pm}
^{\sigma}  = \pm \varsigma \sqrt {a - a_{ \pm} ^{\sigma} }  $ (see
4.2)\,. The derived formulas can be used to construct a
regularized MF (fig.5).

Translation by A.K.~Dambis.

\end{document}